\documentclass[a4paper, 11pt, twoside]{article}
\usepackage{CJK}
\usepackage{latexsym,bm}
\usepackage[tbtags]{amsmath}
\usepackage{amssymb}
\usepackage{fancyhdr}
\usepackage{cite}
\usepackage{tabularx}
\usepackage{booktabs}
\usepackage{dcolumn}
\usepackage{graphicx}
\usepackage{subfigure}
\usepackage{wrapfig}
\usepackage{multicol}
\usepackage{verbatim}
\usepackage{mathrsfs}
\usepackage{multirow}
\usepackage{color}

\usepackage{graphicx}
\usepackage{authblk}

\def\firstpage{1}                           
\newcommand{\supercite}[1]{\!\!\textsuperscript{\cite{#1}}} 

\begin{document} 

\title{{\large  \textbf{Growth and decay of isolated turbulent band in Plane-Couette flow}}
}

\author{\small{Jianzhou LU}}
\author{\small {Jianjun TAO}\thanks{Corresponding author, E-mail: jjtao@pku.edu.cn}}
\author{\small{Weitao ZHOU}} 

\affil{ SKLTCS,  Department of Mechanics and Engineering Science, College of Engineering, Peking University, Beijing 100871, China}
\date{\small{(presented at 25th ICTAM 2020+1. Abstract book, P. 2657-2658.)}  }
\maketitle 
\begin{abstract}
\noindent  The transition of plane Couette flow is numerically investigated in a large computational domain. It is found that the averaged period of the transient growth reduces slowly with the decrease of the Reynolds number (Re) except when Re is close to a threshold value of 286. During the decay process, the band contracts from its both ends with a statistical constant velocity, but keeps its center, width, and tilt angle statistically unchanged. For self-sustained turbulent band, three growth styles are observed. At moderate Re, the isolated band extends obliquely as what happens in plane Poiseuille flow. With the increase of Re, transverse split occurs and the band breaks into several attached segments, forming a shape of `F'. Further increasing Re leads to a longitudinal split, i.e. the isolated band becomes wider at first and then splits into two parallel bands. 
\\[2mm]
\\[2mm]
\\
\end{abstract}


\section{Introduction}
   It is known that plane Couette flow (PCF) is always linearly stable, but may turn to be turbulent at moderate Reynolds numbers by the subcritical transition. Leutheusser and Chu \supercite{Leutheusser} studied the flow between a moving water surface and an upper flat stationary plate, and determined the transitional Reynolds number as 280. It should be noted that the spanwise aspect ratio of the experimental channel was small and the water flow was turbulent, suggesting that the water surface was rough. From then on, substantial progresses have been made and the localized turbulent band or stripe is found to be a key structure for the transition in channel flows \supercite{Tuckerman20}. The band growth and decay processes have been simulated in large domains with under-resolution simulations \supercite{Manneville11, Manneville12}, and it was mentioned that the low resolution lowered the threshold for sustained turbulence from 325 to 210 \supercite{Manneville11} and impeded reliable predictions for fully resolved cases. Therefore, the spatio-temporal evolution of the isolated turbulent bands still requires fully-resolved numerical investigations.

\section{Methods and results}
 We conduct numerical simulations on the PCF in a large domain of the size $800h\times 2h\times 712h$, and a spectral code \supercite{Chevalier07} is used to solve the incompressible Navier-Stokes equations, where the velocity field is expanded in a basis of Fourier modes (in the streamwise $x$- and spanwise $z$-directions) and Chebyshev polynomials (in the wall-normal direction $y$). The numerical resolution is 2048 spectral modes in $x$, 33 in $y$, and 2048 in $z$. The boundary conditions are periodic in the $x$- and $z$-directions and there is no-slip at the walls ($y = \pm h$). The half of the velocity difference between the boundaries U and the half channel gap $h$ are chosen as the characteristic velocity and the length scale, respectively. In the following simulations, an isolated turbulent band, i.e., a unique straight band whose length is several times smaller than the domain size, is used as the initial perturbation.

\begin{figure}[h]
\centering
\includegraphics[scale=0.76]{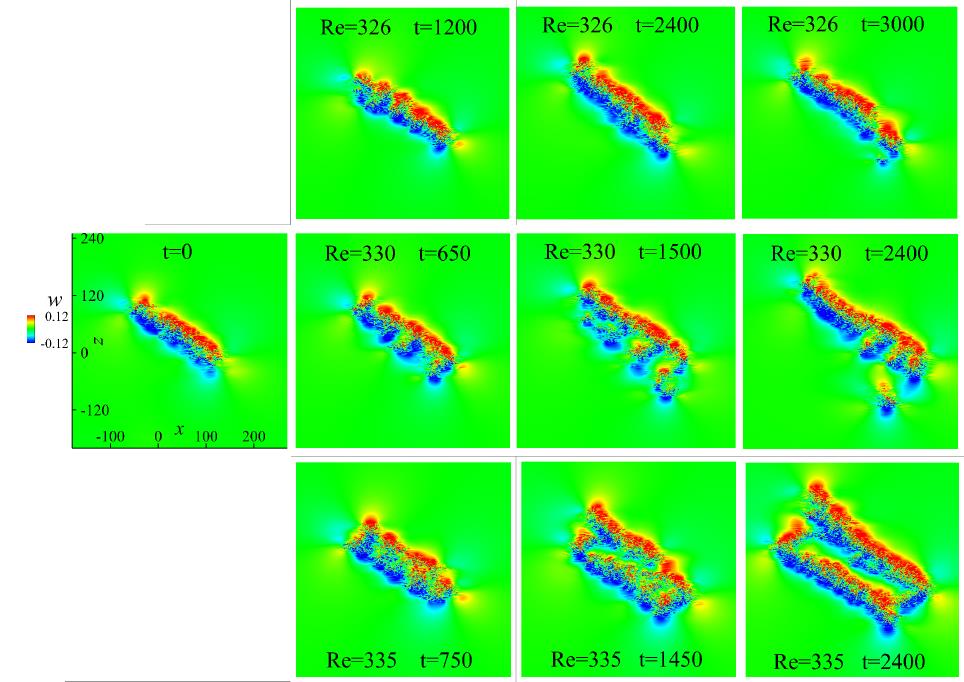}
\caption{ Iso-contours of the spanwise velocity in the mid-plane. The initial fields are the same for all three cases.}
\label{Fig1}
\end{figure}

\begin{figure}[h]
	\centering
	\includegraphics[scale=0.85]{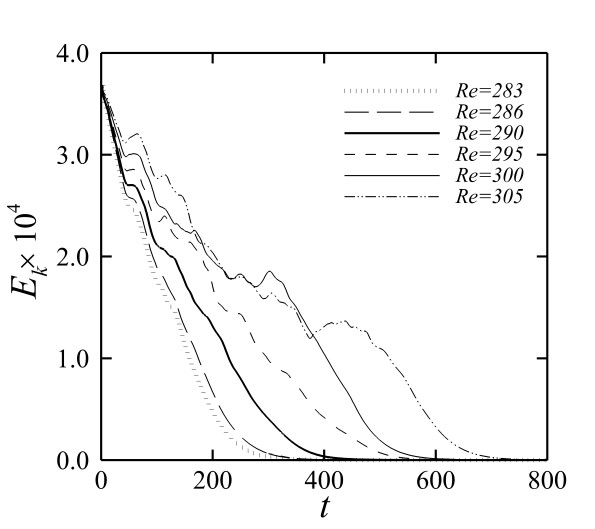}
		\includegraphics[scale=0.85]{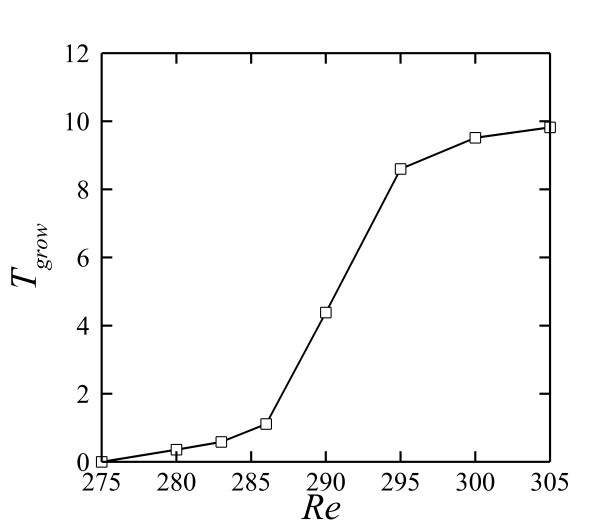}
	\caption{ Time series of the volume-averaged disturbance kinetic energy $E_k$ at different Reynolds numbers obtained with the same initial turbulent band; (b) transient-growth time $T_{grow}$ as a function of Re.}
	\label{Fig2}
\end{figure}

 As shown in Fig. 1, there are three different growth styles for localized turbulence. For moderate Re, e.g. Re=326, the isolated band extends obliquely with time as what occurs in plane Poiseuille flow but with a much lower extension velocity. When Re increases to 330, transverse split occurs and the band breaks into attached segments, forming a shape of `F'. It seems that the transverse split may lead to the lateral branching observed in the previous under-resolution simulations \supercite{Manneville12}. A longitudinal split happens when Re is further increased, e.g. Re=335 as shown in Fig. 1, the isolated band becomes wider at first and then splits into two parallel bands. In fact, the later growth style includes the former one(s), and hence the turbulence spreads more and more efficiently with the increase of Re. 

Since the disturbance kinetic energy of the main body of the isolated band is statistically uniform and the band is straight with a finite width, its shape may be simplified as a tilted rectangle. The center, width, length, and tilt angle of the rectangle can be determined based on the disturbance kinetic energy of the band in the midplane with the method proposed for plane Poiseuille flow \supercite{Tao18}. According to the simulations for $Re=314$, 317, and 318, the position of the band center varies stochastically but does not move much during a period more than 3000 time units, reflecting that the convective velocity of such a localized structure is nearly zero. After an initial period of adjustment, the band contracts longitudinally, i.e. its length decreases generally while its width and tilt angle remain statistically constants. The temporal variation of the band length may be fitted linearly, suggesting a constant contraction velocity. A similar contraction phenomenon was observed before in an under-resolved simulation, where the position of the band center and the shape parameters (e.g., the tilt angle) were not analyzed quantitatively \supercite{Manneville11}. For the present decaying isolated band, the tilt angle is $(27 \pm 3)^{\circ}$ when Re is between 314 and 318.

In order to study the statistical properties of the decay process (Fig. 2a), ten samples are calculated for each Reynolds number, and the initial flow fields are composed by the same band and different random disturbances in the whole fields. We define a parameter, the transient-growth time $T_{grow}$, which is the ensemble average of the mean periods obtained from different samples when the growth rate of $E_k$ is positive, and draw it in Fig. 2(b). It is shown that $T_{grow}$ decreases monotonically by lowering Re, and there is a drastic decline of $T_{grow}$ as Re reduces to 286, indicating a threshold for the transient growth of the turbulent band. When Re<286, the probability of the transient growth is very small, implying that the turbulent band can hardly be formed from the transient growth of any initial disturbances.

\section{Conclusions}
   Direct numerical simulations are carried out to study the subcritical transition of plane Couette flow in a large computational domain. It is found that the isolated turbulent band decays in a style of longitudinal contraction at low Reynolds numbers, and the lower bound Reynolds number of the transient-growth regime is defined as 286. In addition, three different growth styles of the isolated turbulent band are illustrated, i.e. the oblique extension, transverse split, and the longitudinal split. It is believed that their joint action leads to the formation of labyrinthine patterns observed in simulations and experiments.

\end{document}